\documentclass[journal, twocolumn]{IEEEtran}
\usepackage{graphicx}
\usepackage{dblfloatfix}
\usepackage{sidecap}
\usepackage{dcolumn}
\usepackage{bm}
\usepackage[utf8]{inputenc}
\usepackage[T1]{fontenc}
\usepackage{mathptmx}
\usepackage{textcomp}
\usepackage{siunitx}
\usepackage{bigints}
\usepackage{setspace}
\usepackage{hyperref}
\usepackage[switch]{lineno}
\usepackage[table,xcdraw,dvipsnames]{xcolor}
\usepackage{booktabs}
\usepackage{float}
\usepackage{color}
\usepackage{ulem}
\usepackage{multirow}
\begin{document}

\title{Damping Enhancement in YIG at Millikelvin Temperatures due to GGG Substrate }

\author{\IEEEauthorblockN{Rostyslav~O.~Serha$^{1,2}$\IEEEauthorrefmark{1}, Andrey~A.~Voronov$^{1,2}$, David~Schmoll$^{1,2}$,  Rebecca~Klingbeil$^1$, Sebastian~Knauer$^1$, Sabri~Koraltan$^{1,2,3}$, Ekaterina~Pribytova$^4$, Morris~Lindner$^5$, Timmy~Reimann$^5$, Carsten~Dubs$^5$, Claas~Abert$^{1,3}$, Roman~Verba$^6$, Michal~Urbánek$^4$,  Dieter~Suess$^{1,3}$, and Andrii~V.~Chumak$^1$\IEEEauthorrefmark{2}}
\\
\IEEEauthorblockA{
$^1$ Faculty of Physics, University of Vienna, 1090 Vienna, Austria.
\\
$^2$ Vienna Doctoral School in Physics, University of Vienna, 1090 Vienna, Austria.
\\
$^3$ Research Platform MMM Mathematics – Magnetism – Materials, University of Vienna, Vienna, Austria.
\\
$^4$ CEITEC BUT, Brno University of Technology, 61200 Brno, Czech Republic.
\\
$^5$ INNOVENT e.V. Technologieentwicklung, 07745 Jena, Germany.
\\
$^6$ V. G. Baryakhtar Institute of Magnetism of the NAS of Ukraine, Kyiv 03142, Ukraine.
\\
Email: \IEEEauthorrefmark{1}rostyslav.serha@univie.ac.at, \IEEEauthorrefmark{2}andrii.chumak@univie.ac.at}}
\maketitle
%\doublespacing
\begin{abstract}
Quantum magnonics aims to exploit the quantum mechanical properties of magnons for nanoscale quantum information technologies. Ferrimagnetic yttrium iron garnet (YIG), which offers the longest magnon lifetimes, is a key material typically grown on gadolinium gallium garnet (GGG) substrates for structural compatibility. However, the increased magnetic damping in YIG/GGG systems below 50\,K poses a challenge for quantum applications. Here, we study the damping in a 97\,nm-thick YIG film on a 500\,µm-thick GGG substrate at temperatures down to 30\,mK using ferromagnetic resonance (FMR) spectroscopy. We show that the dominant physical mechanism for the observed tenfold increase in FMR linewidth at millikelvin temperatures is the non-uniform bias magnetic field generated by the partially magnetized paramagnetic GGG substrate. Numerical simulations and analytical theory show that the GGG-driven linewidth enhancement can reach up to 6.7 times. In addition, at low temperatures and frequencies above 18\,GHz, the FMR linewidth deviates from the viscous Gilbert-damping model. These results allow the partial elimination of the damping mechanisms attributed to GGG, which is necessary for the advancement of solid-state quantum technologies.
\end{abstract}

\section{Introduction}

Quantum computing promises transformative advances in processing power and problem-solving capabilities, potentially revolutionizing fields such as cryptography, materials science, and artificial intelligence~\cite{Nielsen2010, Arute2019}. In the development of solid-state quantum technology, a key challenge is to identify a viable platform for integrating quantum computing on nanochips, which requires a suitable data carrier capable of operating at the nanoscale~\cite{Chumak2022}. Magnons - the quanta of spin waves - emerge as promising candidates for this role, as they not only cover a wide frequency range from GHz to THz, but also exhibit wavelengths and feature sizes of circuit components down to a few nanometers, typical for modern nanoscale devices~\cite{Chumak2015}.

Quantum magnonics is the field of science that deals with data carried and processed by magnons in magnetically ordered materials at the level of single excitations~\cite{Chumak2022}. 
This field holds promise for breakthroughs in quantum information technologies by facilitating low-dissipation information transfer and enabling coherent magnon-photon, magnon-phonon, and magnon-qubit interactions, which are vital for scalable hybrid quantum networks and devices~\cite{Tabuchi2015, Yuan2022, Li2020}. Furthermore, it offers opportunities in quantum sensing, leveraging the nanoscale magnetic field sensitivity of magnons~\cite{Lachance-Quirion2019}, and expands possibilities through pronounced natural nonlinear and nonreciprocal phenomena~\cite{Chumak2022,Wang2023,Barman2021}.

One of the key challenges in quantum magnonics is the relatively short magnon lifetime compared to the decoherence time of modern superconducting qubits \cite{Krantz2019}. This limitation has driven interest toward materials like yttrium iron garnet (YIG) Y\textsubscript{3}Fe\textsubscript{5}O\textsubscript{12}, renowned for hosting magnons with the longest lifetimes in both bulk form~\cite{SagaYIG, Dillon1957, LeCraw1958, Klingler2017} and thin films grown on gadolinium gallium garnet (GGG) substrates~\cite{Dubs2017, Dubs2020, Ding2020, Heyroth2019, Cornelissen2015, Onbasli2014, Hahn2014}. Although magnon lifetimes remain insufficient for long-term quantum information storage, they are well-suited for enabling the transfer of quantum entanglement between diverse physical systems—including phonons, magnons, and microwave or optical photons—through hybrid quantum magnonic platforms~\cite{Tabuchi2015}. At room temperature, subfields of magnonics focus on leveraging macroscopic quantum states, such as Bose–Einstein condensates (BEC), for data processing~\cite{Demokritov2006, Serga2014, Schneider2020}. Additionally, they explore advanced concepts like magnon phase control by the magnonic Aharonov–Casher effect~\cite{Nakata2017, Savchenko2019, Serha2023}.

To enable operations at the single-magnon level, experiments require millikelvin temperatures to sufficiently suppress the thermal population of phonon and magnon excitations. Single-magnon state operations, in particular, offer access to phenomena such as entanglement, positioning this as a promising frontier in quantum research~\cite{Chumak2022, Yuan2022, Lachance-Quirion2019, Borst2023, Awschalom2021, Babak2022}. Key milestones in the field include the first demonstration of single-magnon control and detection~\cite{Lachance-Quirion2020} and the reconstruction of a single magnon's Wigner function, achieved using a YIG sphere as a magnonic resonator~\cite{Xu2023}. Additionally, YIG has been proposed as a magnonic transducer to link qubits, highlighting its potential for applications in quantum computing~\cite{Li2020}.

To fully harness quantum entanglement in magnonics, transitioning from standing ferromagnetic resonance (FMR) magnons~\cite{Lachance-Quirion2020, Xu2023} to propagating magnons~\cite{Karenovska2018, Knauer2023} is essential. Propagating magnons enable the spatial separation of source and detector, allowing the transport of entangled information across devices—key for scalable quantum technologies. This requires replacing bulk YIG with YIG/GGG thin films, a nanoscale system supporting magnons with lifetimes suitable for coherent information transfer~\cite{Karenovska2018, karenowska2015excitation, Knauer2023}. At low temperatures, however, YIG/GGG systems face about a tenfold  to hundredfold increase in FMR linewidth due to rare earth impurities in the YIG film and the parasitic role of the GGG substrate~\cite{Knauer2023, Guo2022, Jermain2017, Michalceanu2018, Kosen2019}. However, recent work has already shown that by using purer starting materials for growth and by reducing interdiffusion processes at the interface to the substrate, the FMR linewidth of YIG LPE films can be reduced by an order of magnitude compared to physically deposited films at low temperatures~\cite{Cole2023}. The precise mechanisms and their contributions remain largely unexplored, posing challenges for mitigating damping effects required for quantum magnonics.

Building on earlier findings that magnetized paramagnetic GGG substrates generate a stray magnetic field influencing the internal field of YIG films below 50 K~\cite{Danilov1989}, we have recently detailed the formation, magnitude, and pronounced inhomogeneity of this GGG-induced stray field~\cite{Serha2024}.

In this study, we demonstrate that the stray field from a GGG substrate is the dominant contribution to the broadening of the FMR linewidth, a key indicator of magnon lifetime, in a YIG film. Since this broadening is not solely due to the intrinsic magnetic damping of YIG, we refer here to the effective Gilbert damping parameter, as extracted from the experiments. Measurements of FMR linewidth as a function of temperature showed a 13-fold increase in the effective Gilbert damping parameter $\alpha_{\mathrm{eff}}$ and a 5-fold increase in the inhomogeneous linewidth broadening $\Delta B_0$ at frequencies up to 18 GHz compared to room temperature. Specialized numerical simulations and analytical calculations, designed to replicate experimental measurements in the spatially nonuniform GGG stray field, reveal that an increase of a factor up to 6.7 arises from the GGG-induced stray field. Furthermore, above 18 GHz, the linewidth deviates from viscous damping behavior, indicating the limitations of the Gilbert model in describing the system. 

\section{\label{Methods}Methodology}

\subsection{\label{ExpMethods}Experimental methods}

In this work, we studied a (5x5)\,mm$^2$ and 97\,nm-thick YIG film grown in the $\langle111\rangle$ crystallographic direction on a 500\,µm-thick GGG substrate, using liquid phase epitaxy~\cite{Dubs2017, Dubs2020}. We conducted stripline ferromagnetic-resonance (FMR) spectroscopy using a vector network analyzer (VNA), within a Physical Property Measurement System (PPMS), at temperatures ranging from 2\,K to 300\,K and up to 40\,GHz. The sample was mounted on a frequency-broadband stripline and placed in a homogeneous magnetic field of up to 1.3\,T created by superconducting coils. The maximum applied microwave power in the PPMS was -5\,dBm. For temperatures below 2\,K, measurements were conducted in a dilution refrigerator, achieving base temperatures around 10\,mK. At 20\,mK, the refrigerator provides a cooling power of 14\,µW, which is sufficient to maintain thermal equilibrium during FMR spectroscopy measurements with an applied power of -25\,dBm.

The following measurements were performed with the external magnetic field in the in-plane $\langle1\overline{1}0\rangle$ crystallographic direction and applied along the FMR stripline antenna (see Fig.~\ref{f:1}~(a)). To obtain the FMR spectrum at a specific field, the transmission parameter $S_{12}$ was measured using a VNA. Measurements were performed not only at the target field but also at reference fields, offset by approximately 15\,mT to 40\,mT above and below the target value~\cite{Weiler2018}. By subtracting the averaged signals of the reference fields from the measured FMR signal, we minimized the magnetically active background signal originating from the GGG substrate, obtaining the FMR absorption spectrum only in YIG (see Fig.~\ref{f:1}~(c) as an example). This dual reference measurement approach enabled to obtain the best results when working with kelvin and sub-kelvin temperatures, since the GGG parasitic offset signal is greatly affected by the change in the applied field. To obtain the resonance frequency and full linewidth at half maximum (FWHM), the background is first analyzed using a 1D cubic spline model~\cite{herrera2023double}. The resonance shape is then fitted using the split-Lorentzian model, which individually describes the left and right sides of the asymmetric absorption peaks.

To accurately determine the magnetization of GGG for our analytic calculations and numerical simulations, we utilized vibrating-sample magnetometry (VSM) on a pure GGG slab in the temperature range from 1.8\,K to 300\,K. The raw measurement VSM data is dependent on the sample shape due to self-demagnetization, which has to be recalculated into the true material law to use in numerical simulations~\cite{Serha2024}. The Gd\textsuperscript{+3} ions in GGG have a relatively large spin (S\,=\,7/2), resulting in a saturation magnetization $M_{\textup{GGG}}^{\textup{s}}$ = 805\,kA/m, which is notably higher than that of YIG.

By linearly fitting the FWHM vs the FMR frequency we obtain the effective magnetic damping parameters, effective Gilbert damping parameter $\alpha_{\mathrm{eff}}$ and the inhomogeneous linewidth broadening $\Delta B_{0}$ (see~\cite{Boettcher2022} and the supplementary materials therein), of a magnetic material. In this study, the temperature-dependent saturation magnetization of YIG, $M_{\textup{YIG}}^{\textup{s}}$, is taken from the analytical calculation performed in~\cite{Hansen1974} and the anisotropy fields are taken from~\cite{Serha2024}.  

As reported in the literature~\cite{Knauer2023, Danilov1989} and supported by our FMR analysis, the paramagnetic GGG becomes sufficiently magnetized at temperatures below about 100\,K, together with an external magnetic field applied. This magnetization induces a magnetic stray field $B_{\textup{GGG}}$ in the YIG layer, which causes a shift of the YIG FMR frequencies~\cite{Danilov1989,Serha2024}. For the in-plane applied magnetic field, the FMR shift is toward lower frequencies because $B_{\textup{GGG}}$ and the applied bias field $B_{\textup{0}}$ are antiparallel. Conversely, in an out-of-plane geometry, the stray field $B_{\textup{GGG}}$ aligns parallel to the field $B_{\textup{0}}$, resulting in a shift of the resonance frequency to higher values~\cite{Danilov1989}. The magnitude of this inhomogeneous stray field $B_{\textup{GGG}}$ is influenced by both the temperature and the strength of the external magnetic field. At lower temperatures and higher external fields, the GGG-induced stray field becomes more pronounced.

\subsection{\label{NumMethods}Numerical simulations}

To gain a better understanding of the magnetic stray field present in the YIG layer on top of the GGG substrate, we performed numerical simulations to provide a more precise representation of the field inhomogeneity compared to analytical calculations.

\begin{figure*}[b!]
    \includegraphics[width=1\linewidth]{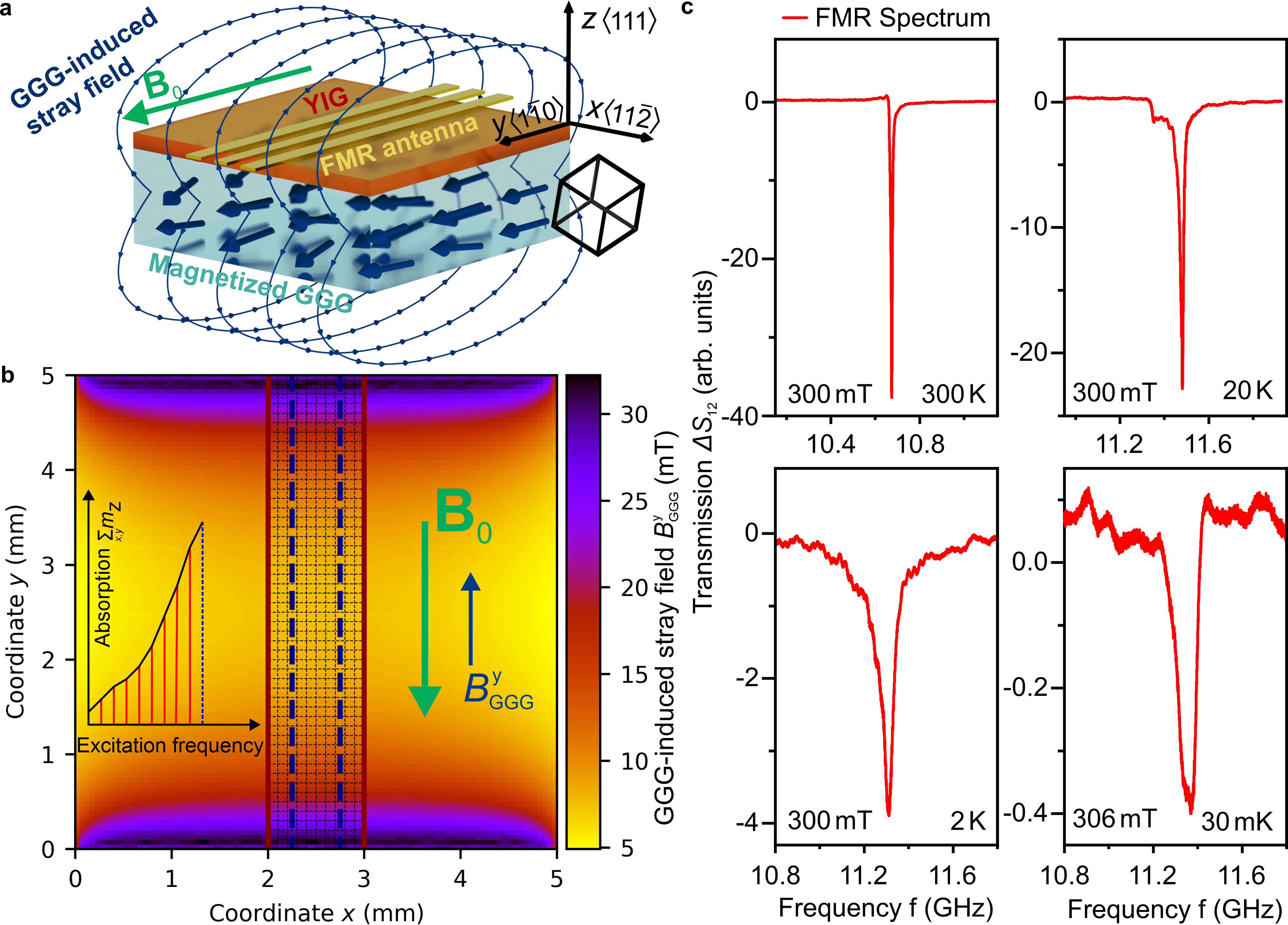}
    \caption{(a) Depiction of the experimental FMR system of a YIG film grown on GGG. The crystallographic directions are indicated along the axes, and the orientation of the cubic lattice is shown by the black frame cube. The sample is in-plane magnetized along the $\langle1\overline{1}0\rangle$ direction by an external magnetic field. At temperatures approaching 0\,K the paramagnetic GGG spin system saturates. The substrate creates an inhomogeneous stray field, that becomes an additional component to the internal magnetic field of the YIG but is oriented antiparallel to the external field. To perform FMR measurements, the YIG film is placed on a CPW antenna, which excites the magnetic system.
    (b) Simulation of the highly inhomogeneous GGG-induced stray field y-component $B_{\textup{GGG}}^\textup{y}$ at the interface between YIG and GGG layers. The inset shows schematically the concept of micromagnetic simulations: the FMR absorption is simulated as the sum of the magnetization components for all cells within the excitation region for each applied frequency. The red vertical lines show the geometry used for the FMR micromagnetic simulations with the dashed blue lines indicating the excitation region.
    (c) Example measured spectra of FMR at the temperatures of 300\,K, 20\,K, 2\,K, and 30\,mK for an external magnetic field of 300\,mT.}
    \label{f:1}
\end{figure*}

The simulations were conducted using the finite-difference micromagnetic solver \texttt{magnum.np}~\cite{bruckner2023}. However, since the YIG film used in the experimental studies is several orders of magnitude thinner than the GGG substrate, simulating both materials together while maintaining the unified discretization required by the finite-difference method is not feasible. Consequently, the GGG-induced stray field was computed separately using the FEMME software, which employs the finite element / boundary element method to solve the magnetostatic Maxwell equations in the GGG substrate region. Nonlinear magnetization curves, corresponding to various studied temperatures, were provided as input for the calculations~\cite{Serha2024, Dieter2012}. The resulting field was evaluated 50\,nm above the GGG surface, within the area of the YIG film, as shown in Fig.\,\ref{f:1}\,(b).

This computed stray field was then used as a static bias field, $B_\mathrm{GGG}$, in subsequent micromagnetic simulations to investigate its influence on the FMR in the YIG film. To properly capture the FMR phenomenon, we performed multiple simulations at different excitation frequencies $f$ while applying the same external bias field, $B_0$. For each frequency, the microwave power $P(f)$ absorbed by the magnetic film was calculated as the scalar product of the space- and time-dependent magnetization $\mathbf{m}(\mathbf{r}, t)$ of the YIG and the oscillating excitation field $\mathbf{b}(\mathbf{r},f,t)$~\cite{gurevich1963}:

\begin{equation}
\label{eq:absorp_integral}
P(f) \sim f \cdot \mathrm{Im} \left[\int_V\mathrm{d}^3 \mathbf{r} \frac{1}{T} \int_t \mathrm{d}t \left(\mathbf{m}(\mathbf{r}, t)\cdot\mathbf{b}(\mathbf{r},f,t)\right)\right].
\end{equation}

In the simulations, the excitation field was uniform in space and oriented perpendicular to the plane of the film, defined as $\mathbf{b}=\{0,0,b_\textup{z}\cos{(2\pi ft)}\}$. Under these assumptions, the integrals in Eq.~\eqref{eq:absorp_integral} were replaced by a summation over time for numerical calculations:

\begin{equation}
\label{eq:absorp_sum}
P(f) \sim f^2\cdot\mathrm{Im}\left[\sum_{t_j=t_0}^{t_0+T}m_\mathrm{z}(t_j)\mathrm{exp}({2\pi i ft_j)}\Delta t_j\right],
\end{equation}

where $m_\textup{z}$ is the spatially averaged z-component of the unit magnetization vector, $t_0=2/(\alpha f)$ is the minimum simulation time required for the magnetization to reach steady oscillations, $T=1/f$ is the oscillation period, and $\alpha$ is the Gilbert damping parameter for YIG. According to Eq.\,\eqref{eq:absorp_sum}, a sufficiently long simulation time $t_0$ is required to compute the absorption for each frequency point in the FMR spectrum.

To reduce computation time, we simulated a smaller $1\times\nolinebreak5\,$mm$^2$ section of the YIG film, with a discretization cell size of $5\times5\times0.1$\,µm$^3$, as indicated by the red vertical lines in Fig.\,\ref{f:1}\,(b). This region included the $0.5\,$mm-broad excitation area, where the oscillating field was applied (blue dashed lines in Fig.\,\ref{f:1}\,(b)), and where the averaged magnetization $m_\mathrm{z}$ was computed. To eliminate the influence of spin-wave back reflections on $m_\mathrm{z}$, the Gilbert damping parameter $\alpha$ was exponentially increased near the borders of the simulation mesh outside the excitation region, as spin-wave propagation was not the focus of this study.

Finally, the simulated absorption spectra were fitted to determine the FWHM using an asymmetrically shaped Split-Lorentzian function implemented in the curve-fitting package \texttt{lmfit}~\cite{lmfit}.

\subsection{Semianalytical model}\label{s:theory}

For theoretical calculations of FMR curves, we use the model of independent resonating areas, subjected to different local magnetic field due to GGG stray field. This model is applicable if all the cooperative phenomena, i.e. spin wave propagation and/or formation of standing spin-wave modes, are suppressed by the dissipation on the scale of the sample and field nonuniformity. This assumption fits well our case, as the lateral sizes of YIG (5\,mm) are much larger than the spin-wave mean free path (which is less than 100\,µm in the parallel to the field and antenna direction even at room temperature due to nanoscale YIG thickness). 

The local FMR angular frequency in our geometry of in-plane magnetized film is given by
\begin{equation}
	\omega(B_\mathrm{loc}) = \gamma \sqrt{B_\mathrm{loc}\left(B_\mathrm{loc} + B_\mathrm{eff}^\mathrm{(an)} + \mu_0 M_\mathrm{YIG}^\mathrm{s} \right)} \,,
\end{equation}
where the temperature-dependent YIG saturation magnetization $M_\mathrm{YIG}^\mathrm{s}$ and effective anisotropy $B_\mathrm{eff}^\mathrm{(an)}$ are included, see \cite{Serha2024} for details. The local field $B_\mathrm{loc}$ is given by the external one $B_0$ and GGG stray fields, and is expressed as
\begin{equation}
	B_\mathrm{loc}(y) = B_0 - \mu_0 M_\mathrm{GGG} N_{yy}(y) \,,
\end{equation}
with $N_{yy}(y)$ being the $yy$-component of the coordinate-dependent mutual demagnetization tensor of YIG/GGG bilayer. Accounting for small YIG film thickness, we can use a simplified expression for the mutual demagnetization tensor of $2a\times 2a\times 2c$ prism at its surface  \cite{Smith2010, Joseph1965}
\begin{equation}
	N_{yy}(y) = \frac1{2\pi} \left(F[y+a] - F[y-a] \right) \,,
\end{equation} 
where we define
\begin{equation}
	F[Y] \equiv \arctan \frac{a}{Y\sqrt{a^2 + Y^2 + 4c^2}} \,,
\end{equation}	
and assume center position in the $x$-direction (i.e., under the antenna).
The power absorbed by a local region at the excitation frequency $\omega_\mathrm{e}$ is proportional to \cite{Gurevich1996} $P \sim \mathrm{Im} [(\omega_{\mathrm{loc}} + i\Gamma - \omega_\mathrm{e})^{-1}]$, where $\Gamma$ is the damping rate. Then, the total absorption, i.e., the resonance curve, is given by the integration:
\begin{equation}\label{e:P-f-fin}
	P(B_0,\omega_\mathrm{e}) \sim \mathrm{Im}\left[ \int\limits_{-a}^a \frac{1}{\omega (B_\mathrm{loc}(y) + i\Delta B) - \omega_\mathrm{e}} dy\right] \ .
\end{equation}
Here we account for the inherent YIG damping by adding imaginary part of the effective field $\Delta B$, which was extracted from room-temperature FMR measurements. In a general case, there is an additional multiplier under the integral, which is related to the field-dependent precession ellipticity, but its effect is vanishing until spatial variations of the field are much less than the mean field value.
  
\section{Results}

Fig.\,\ref{f:1}\,(c) displays four spectra that exemplify the impact of cryogenic temperatures on the FMR signal of the YIG film in an external magnetic field of 300\,mT. The FMR peak is almost entirely Lorentzian-shaped at 300\,K. However, when the temperature decreases below 100\,K, the peak becomes asymmetric on one side, here to the side of lower frequencies, as shown in Fig.\,\ref{f:1}\,(c) for 20\,K, and then broadens significantly with a lower resonance frequency, as demonstrated for 2\,K and 30\,mK. Additionally, the amplitude of the FMR peak decreases as the temperature reduces. This asymmetric broadening, which can increase by up to a factor of ten, becomes clear when considering the distribution of the GGG stray field~$B_{\textup{GGG}}^\textup{y}$. It reduces the internal field of YIG at the edge more than at the center of the film, thus, the same happens to the FMR frequency~$f_{\mathrm{FMR}}$, asymmetrically broadening the FMR peak towards lower frequencies. The dependence of the resonance frequency $f_{\mathrm{FMR}}$ on the GGG magnetization is discussed in detail in~\cite{Serha2024}. It is to be noted that this effect is expected to be more pronounced in systems where the spin-wave propagation length is shorter than the sample size.

\begin{figure}[!t]
    \includegraphics[width=0.97\linewidth]{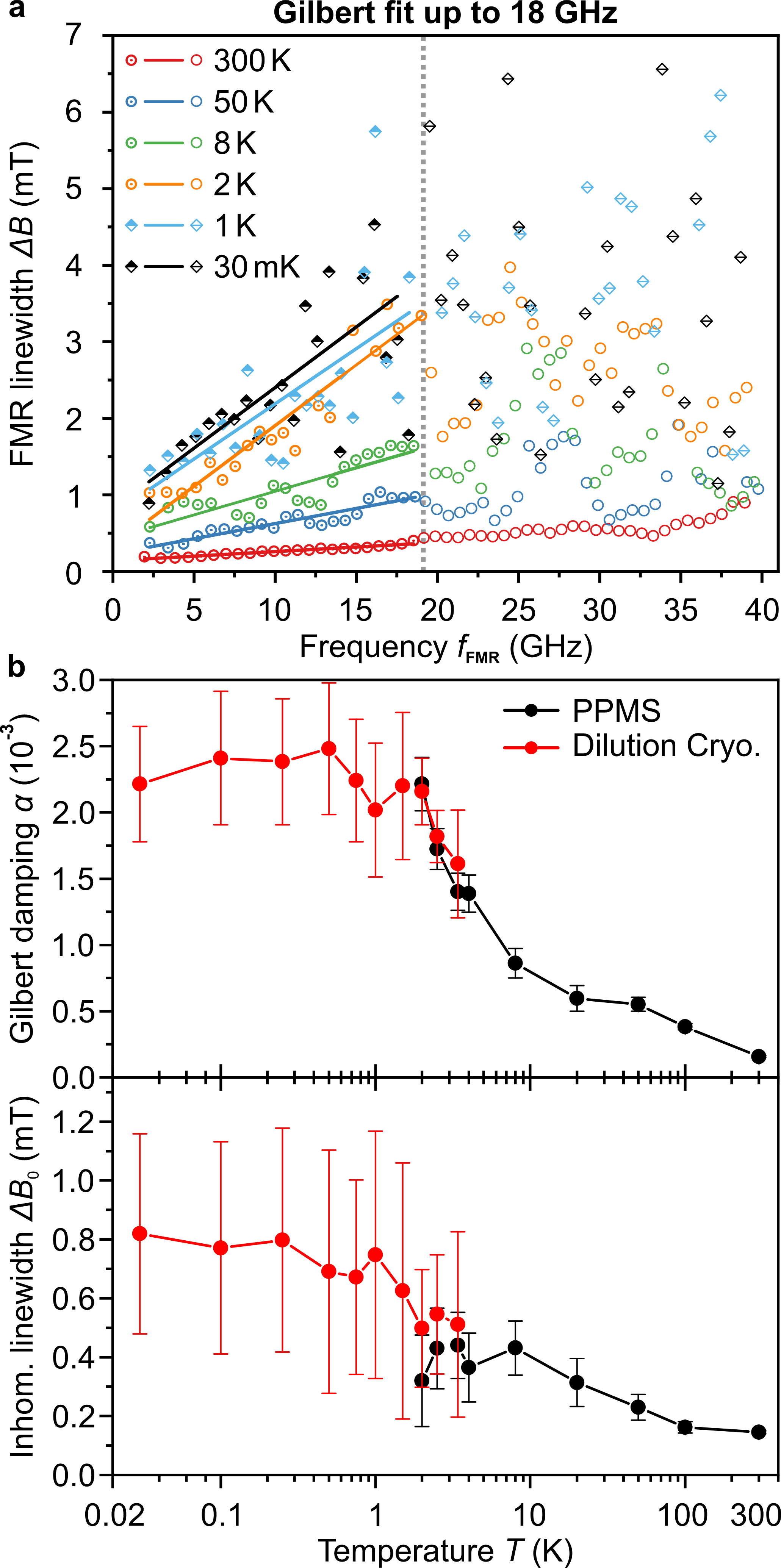}
    \caption{(a) Experimental FMR linewidth $\Delta B$ as a function of the FMR resonance frequency $f_{\mathrm{FMR}}$. The round points are measured in the PPMS setup, while the diamond points are taken from measurements in the dilution refrigerator. The half-solid points are measurement points up to 18\,GHz (gray dashed line), the hollow points are measurements above 18\,GHz and the straight lines are Gilbert fits performed up to 18\,GHz. Above this value the linewidth portrays a non Gilbert behavior vs $f_{\mathrm{FMR}}$. (b) Effective Gilbert damping parameter $\alpha_{\mathrm{eff}}$ and inhomogeneous linewidth $\Delta B_{0}$ vs the temperature $T$ in a $x$-axis logarithmic scale for FMR frequencies up to 18\,GHz. The fits were performed for measurements taken from the PPMS setup (black) and the dilution cryostat (red).}\label{f:2}
\end{figure}

Fig.\,\ref{f:2}\,(a) shows the impact of the stray field induced by the GGG substrate. The graph displays the FMR linewidth $\Delta B$ vs the FMR frequency $f_{\mathrm{FMR}}$ obtained for the YIG film at six different temperatures, ranging from 300\,K to 30\,mK. For 300\,K the linewidth behavior is following the linear Gilbert damping model. However, as the temperature decreases, the FMR linewidth increases. Up to the frequency of approximately 18\,GHz the linewidth shows linear behavior and can be fitted with the Gilbert model portrayed by the solid lines. However, above 18\,GHz the linear Gilbert model is not applicable anymore, the linewidths are highly scattered and do not portray a linear increase with FMR frequency at higher fields. Apparently at some high FMR frequencies the linewidths become as narrow as at low FMR frequencies. This is clearly visible in Fig.\,\ref{f:2}\,(a) for 8\,K shown in green, where the FMR linewidth at around 40\,GHz has the same value as at 5\,GHz. In general, a non-Gilbert-like behavior is expected in this system, since increasing the field not only increases the FMR linewidth as predicted by the Gilbert model, but also increases the magnetization of the GGG substrate, which is responsible for the parasitic FMR broadening. However, this behavior should exhibit a monotonic trend and cannot explain the "oscillatory" behavior observed in the experiment. The exact physical mechanisms underlying our observations require further investigation, ideally using alternative methods to determine magnetic damping or magnon lifetime in YIG.

To analyze the performed Gilbert fits in Fig.\,\ref{f:2}\,(a), the two fit parameters, effective Gilbert damping $\alpha_{\mathrm{eff}}$ and inhomogeneous linewidth broadening $\Delta B_{0}$, are shown vs temperature in Fig.\,\ref{f:2}\,(b) and (c). At 300\,K the values for $\alpha_{\mathrm{eff}}$ and $\Delta B_{0}$ are $1.6\times10^{-4}$ and 0.145\,mT accordingly, stating the very high quality of the YIG film. But with decreasing temperature the parameters worsen as the effective Gilbert damping $\alpha_{\mathrm{eff}}$ and the inhomogeneous linewidth broadening $\Delta B_{0}$ are both increasing up to a factor of 13 and 5 accordingly.
Around a temperature of around 500\,mK the increase of both the Gilbert fit parameters flattens out, which corresponds with the saturation of the magnetization behavior of the paramagnetic GGG with lowering the temperature, which was described in~\cite{Serha2024,Deen2015}. This behavior is hinting that the increase in linewidth at low temperatures has one of its origins in the magnetization of the GGG substrate.

As the measurements were performed within two different setups, the PPMS and dilution cryostat, it is worth noting that the results of overlapping temperatures are within each others error bars, proving consistency of the measurements. The error bars are taken from the root-mean-square deviation of the fit. The measurements within the dilution refrigerator exhibited a lower signal-to-noise ratio, which led to a higher scattering of the data points in Fig.\,\ref{f:2}\,(a) and larger error bars in Fig.\,\ref{f:2}\,(b) and (c) in red. It was necessary to maintain the applied power at a low level (below -25\,dBm) to ensure that the system remained in thermal equilibrium, which resulted in a more noisy signal (see Sec.\ref{ExpMethods}). 

Several effects are known from literature that can be attributed to the broadening of the FMR linewidth in YIG/GGG systems. The first and well studied effect in bulk YIG and YIG films is due to relaxation mechanisms of rare earth impurities in GGG and YIG~\cite{Michalceanu2018, Dillon1959, Spencer1959, Sparks1961,Seiden1964,Boventer2018, Schmoll2024}. The second potential mechanism is the dipolar coupling of the YIG system with the partially magnetized GGG substrate, which inherently has a large EPR linewidth of approximately 400\,mT~\cite{Barak1992,Bedyukh1999}. However, as recently shown in ~\cite{ Schmoll2024}, this mechanism is pronounced for propagating magnons with non-zero wavenumbers $k \neq 0$ and should vanish for FMR as well as for short-wavelength exchange magnons. The third mechanism is related to the GGG stray field and is studied in this paper. 

To fully understand the experimental FMR results, it is essential to examine the nature of the GGG-induced stray field, which can be most precisely explored through simulation. Determining the profile and strength of this stray field within the YIG layer required characterizing the magnetization behavior of the bare GGG substrate, as outlined in \cite{Serha2024}. Even in the partially magnetized state shown in Fig.\,\ref{f:1}\,(a), GGG generates a highly inhomogeneous $y$-component stray field\, $B_{\textup{GGG}}^\textup{y}$ within the YIG film, as illustrated in Fig.\,\ref{f:1}\,(b). At 2\,K, this induced field\, $B_{\textup{GGG}}^\textup{y}$ opposes the external magnetic field of 300\,mT and varies in magnitude from 8\,mT in the center to the maximum of 33\,mT at the edges. This stray field significantly impacts the investigation of YIG/GGG systems at low temperatures, particularly by parasitically broadening the FMR peak. Since this effect depends on the magnetization of the GGG substrate, which varies across the external field range, it influences both the effective Gilbert damping coefficient $\alpha$ and the inhomogeneous linewidth broadening $\Delta B_0$.

\begin{figure}[p!]
    \includegraphics[width=0.95\linewidth]{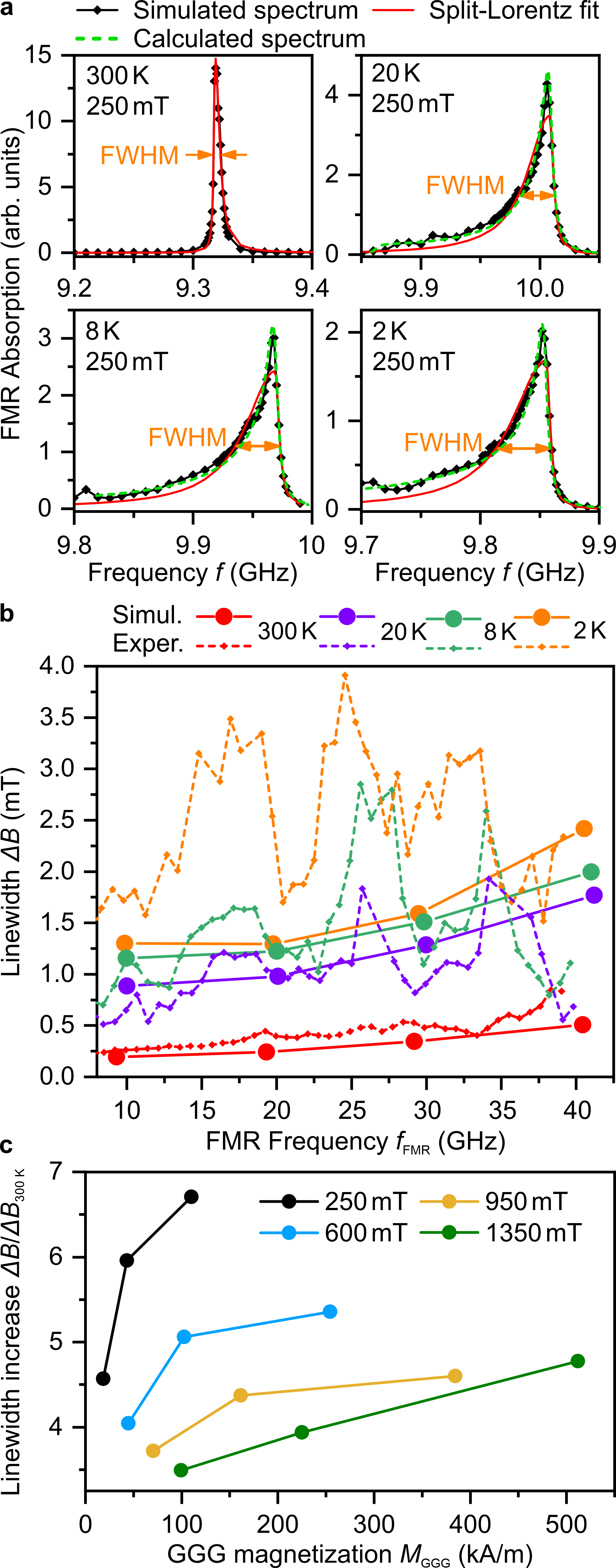}
    \caption{(a) Example of simulated FMR spectra (black) with split-Lorentz fit (red) at 250\,mT for temperatures of 300\,K, 20\,K, 8\,K, and 2\,K. For the latter three temperatures, the graphs also include analytical calculations represented by dashed green lines. With lowering the temperature the FMR spectra become asymmetric broadened to lower frequencies and, thus, their linewidth (FWHM) increase. (b) Simulated FMR linewidth vs FMR frequency for four temperatures compared to experimental data. The Simulated results are depicted with round dots and connected with lines for visual guidance, while the experimental data are represented by dashed lines and diamond-shaped dots for the measurement points. (c) Increase factor $\Delta B / \Delta B_{\textup{300\,K}}$ of the simulated linewidth at low temperatures in comparison to 300\,K vs the GGG magnetization $M_{\textup{GGG}}$. Points are connected by lines to aid visual guidance.}\label{f:3}
\end{figure}

To extract the linewidth increase caused by the GGG nonuniform stray field, simulations of the FMR absorption of YIG in the biasing GGG-induced stray field were performed (see Sec.\,\ref{NumMethods}). In Fig.\,\ref{f:3}\,(a), four simulated spectra (black) are shown for different temperatures at 250\,mT. The simulated FMR peaks were fitted (red line) using the same methods as the experimental data to better extract and compare the FWHM with the experiment. With higher magnetization of the GGG, which occurs at lower temperatures, the simulated FMR becomes broader, the absorption amplitude decreases, and the peaks show clear asymmetry towards lower frequencies compared to the 300\,K simulation, in which the GGG magnetization is negligible. This behavior is similar to the experimental results shown in Fig.\,\ref{f:1}\,(c), confirming the asymmetrical parasitic broadening of the FMR linewidths by the GGG-induced stray field.

In addition, in Fig.\,\ref{f:3}\,(a) we show theoretically calculated with Eq.\,(\ref{e:P-f-fin}) FMR curves. For the calculations we set the inherent YIG damping as $\Delta B_0[\mathrm{mT}] \approx 0.06 + 0.006 f$, where the frequency $f$ is in GHz units, obtained from room-temperature measurements. One finds very good agreement of the calculated and simulated curves, with some small deviations caused, most probably, by the interaction of resonating areas, which is neglected in the theoretical model. In the following, we show analysis of micromagnetic data, the difference with the characteristics extracted from analytically calculated curves is impossible to distinguish on the used data scales. 

The full set of simulations is compared to the corresponding experimental datasets in Fig.\,\ref{f:3}\,(b), where the FMR linewidth $\Delta B$ is plotted against the FMR frequency~$f_{\mathrm{FMR}}$. Due to the significant time and computational resources required, we conducted simulations at four temperatures, each with four different external magnetic fields. At 300\,K (red), the simulations exhibit a clear linear Gilbert damping behavior and closely match the experimental results, confirming their overall accuracy and resemblance to the true linewidths in YIG films. At lower temperatures, the linewidths increase for both simulations and experimental data. At 2\,K, the simulated linewidth reaches 2.42\,mT for the maximum GGG magnetization value available in the experiment. The linewidths for 20\,K (violet) and 8\,K (green) are roughly consistent in linewidth value between simulations and experiments, within the scattering of the experimental data. A discrepancy is observed at high frequencies around 40\,GHz, where the experimental FMR spectra show significantly lower linewidths, much lower than the simulations. At even lower temperatures, approaching 2\,K (orange), the experimentally obtained linewidths are significantly higher, reaching up to 4\,mT, compared to the simulated values. This deviation suggests that the broadening caused by the inhomogeneous GGG stray field is not the only mechanism at play, but additional processes are contributing to increased effective damping in YIG/GGG systems at low temperatures.

To quantitatively describe the simulated linewidth increase caused by the inhomogeneous GGG-induced field, the linewidths at low temperatures are normalized with respect to the 300\,K linewidths at the corresponding magnetic field and plotted against the magnetization of the GGG substrate~$M_{\textup{GGG}}$ in Fig.\,\ref{f:3}\,(c). It is evident that at low external fields~$B_{0}$ and FMR frequencies~$f_{\mathrm{FMR}}$, the increase in linewidth due to the GGG-induced stray field is maximal, reaching a factor of 6.7 at 2\,K and a field of 250\,mT. In this particular point the experimental increase in the FMR linewidth compared to room temperature is about 6.6 and is comparable to the numerical value. The lowest increase in the simulated FMR linewidth is observed at 20\,K and 1350\,mT, still reaching 3.5 times. This behavior of the linewidth increase $\Delta B/\Delta B_{\mathrm{300\,K}}$ clearly shows that the increase of the GGG stray field is the dominant mechanism responsible for the FMR linewidth broadening in the sample under investigation. It has particularly high impact for the cases of narrow linewidths at low external fields and FMR frequencies in contrast to high external fields and FMR frequencies. From Fig.\,\ref{f:3}\,(c), it is clear that to measure the true FMR linewidths, the parasitic effect of linewidth broadening must be eliminated across the entire range of external magnetic fields at low temperatures, particularly at low magnetic fields. This can be accomplished by reducing the gradient of the GGG-induced stray field over the FMR measurement area, achievable through microstructuring of the YIG film or by optimizing the substrate shape and magnetization geometry.

\section{Potential materials for quantum magnonics}

Despite the challenges of YIG/GGG systems at low temperatures, YIG remains the leading material for studying quantum magnonics. Advances in scaling YIG structures to sub-micrometer dimensions have been driven by the development of high-quality, nanometer-thick YIG films fabricated using LPE~\cite{Dubs2020, Pirro2014, Hahn2013, Althammer2013}, pulsed laser deposition (PLD)~\cite{Onbasli2014, Sun2012, Yu2014}, and sputtering~\cite{Liu2014,Schmidt2020,Ding2020} on GGG, achieving thicknesses as small as sub-10\,nm.

The integration of YIG films with yttrium aluminum garnet (YAG) substrates aims to unlock new possibilities for quantum magnonics by replacing the paramagnetic GGG substrate with a diamagnetic YAG substrate, particularly interesting for cryogenic applications. However, despite the drawbacks of GGG at low temperatures, the quality of YIG/GGG systems still surpasses that of YIG/YAG systems \cite{Sposito2013,Krysztofik2021} due to the large lattice mismatch between YIG and YAG. Other materials have shown greater promise as potential diamagnetic substrates for YIG. For instance, yttrium scandium aluminum garnet (YSAG), used as a diamagnetic spacer between YIG and GGG, has demonstrated reduced damping in YIG films at low temperatures \cite{Guo2022}. However, the damping values achieved at low temperatures remain considerably higher than those observed in high-quality YIG LPE films~\cite{Cole2023}. In general, the broad garnet family \cite{Yoshimoto2019} holds potential for utilizing other diamagnetic substrates with lattice constants closely matching YIG.

Beyond YIG, other materials are emerging as promising candidates for quantum magnonic experiments and applications~\cite{Kruglyak2024}. One example is the van der Waals antiferromagnet CrPS$_4$, which exhibits long-distance magnon transport~\cite{Wal2023}. Similarly, van der Waals magnetic materials such as CrCl$_3$ have demonstrated standing spin waves, with experiments observing standing spin-wave modes across a thickness of 20\,µm, highlighting the potential for manipulating spin-wave modes in 2D magnetic systems~\cite{Kapoor2020}. Furthermore, propagating spin waves were detected in the 2D van der Waals ferromagnet Fe$_5$GeTe$_2$, despite a relatively high magnetic damping of such materials~\cite{Schulz2023}. These findings underscore the potential of 2D materials to advance magnonic technologies, offering tuneable and versatile platforms for future magnonic applications.

Organic materials also present exciting possibilities due to their tunable magnetic properties and potential for low-damping spin-wave propagation. For instance, the organic-based magnet V(TCNE)$_x$ has exhibited extremely low magnetic damping comparable to YIG~\cite{Liu2020, McCullian2020}. This makes organic materials particularly attractive for hybrid quantum magnonic systems, where their flexibility and chemical versatility could enable integration with superconducting circuits or other quantum platforms. Furthermore, the ability to chemically tailor these materials provides a pathway for engineering magnon lifetimes and coupling strengths.

Metallic compounds such as CoFeB ~\cite{Talmelli2020, Wojewoda2024} and CoFe ~\cite{Schoen2016}, which are also used in applied magnonics, can also be utilized for quantum magnonics due to their versatility in being deposited on different substrates. Their compatibility with existing spintronic technologies and scalability in fabrication processes further enhance their attractiveness. However, their magnon lifetimes are expected to be shorter than those of YIG.

\section{Conclusion}
We show that the FMR linewidth of YIG films grown on GGG substrates is significantly affected by stray fields originating from the partially magnetized paramagnetic GGG at low temperatures under externally applied magnetic fields. The strength and configuration of these stray fields depend on the geometric shape of the GGG substrate (the ratio of its width, length, and thickness) and exhibit strong inhomogeneity across the YIG layer. In an in-plane magnetization geometry, micromagnetic simulations and semianalytical calculations showed that these stray fields can increase the FMR linewidth by up to 6.7 times, the value which matches well also the experimental findings. The anomalous temperature-independent effective damping behavior of YIG below 500\,mK, is attributed to the absence of variation in the GGG magnetization $M_{\textup{GGG}}$ with decreasing temperature, and consequently, to the GGG-induced magnetic field $B_{\textup{GGG}}$. This behavior reflects the properties of GGG as a geometrically highly frustrated magnet, characterized by its complex phase transition diagram at low temperatures and applied fields~\cite{Schiffer1994, Tsui1999, Petrenko1998, Deen2015, Serha2024}.

By engineering the dimensions of the YIG layer and the GGG substrate, it is possible to create conditions where the YIG device or network resides within a relatively uniform GGG stray field, enabling efficient magnon transport. However, complete elimination of the adverse effects caused by the GGG substrate can only be achieved by replacing GGG with a diamagnetic substrate capable of ensuring the same quality of YIG as grown on GGG.

In summary, our findings enable precise predictions of the magnetic behavior of YIG/GGG systems at low and ultra-low temperatures, which is crucial for the successful development of future YIG/GGG quantum-magnonic networks. Furthermore, these phenomena are broadly applicable to any magnetic thin film grown on a paramagnetic substrate at low temperatures under an applied magnetic field.

\section*{acknowledgements}
This research was funded in part by the Austrian Science Fund (FWF) project Paramagnonics [10.55776/I6568].
SK acknowledges the support by the H2020-MSCA-IF under Grant No. 101025758 (“OMNI”).
RV acknowledges support of the NAS of Ukarine (projects 0123U104827 and 0123U100898).
The work of ML was supported by the German Bundesministerium für Wirtschaft und Energie (BMWi) under Grant No. 49MF180119. CD thanks R. Meyer (INNOVENT e.V.) for technical support.

\section*{Author contributions}
ROS conducted all measurements, processed and analyzed the data, and authored the initial draft of the manuscript.
AAV~executed the micromagnetic simulations and contributed to data analysis.
AAV and RK have implemented the data fitting method.
SAK, CA and DS developed the software used for micromagnetic simulations.
DAS, SAK, and SK constructed the experimental setup and supported the experimental investigations.
MU and EP assisted in interpreting the experimental data and provided insights into the measurement results.
ML, TR and CD synthesized the YIG film. 
RV provided theoretical support.
AVC~planned the experiment and led the project.
All authors discussed results and contributed to the manuscript.

\section*{Competing interests}
The authors declare no competing interests.
%\section*{Additional information}

%\textbf{Supplementary information} is available for this paper at .

%\textbf{Reprints and permission information} is available at .
\textbf{Data availability}
The data that support the findings of this study are available
from the corresponding author upon reasonable request.
\emergencystretch=5em
\bibliographystyle{BST}
\bibliography{BIB}

%\printbibliography

\end{document}